\documentclass{osa-article}

\journal{osajournal}

\articletype{Research Article}

\usepackage{lineno}
\usepackage{amsmath}
\usepackage{graphics}
\usepackage{xcolor}
\usepackage{threeparttable}
\usepackage{algorithm}
\usepackage{algpseudocode}
\usepackage{upgreek}
\modulolinenumbers[5]
\usepackage{multirow}
\usepackage{array}

\begin{document}

\title{Controllable energy angular spectrum method}

\author{Fan Wang,\authormark{1,*} Tomoyoshi Shimobaba,\authormark{1} and Takashi Kakue\authormark{1},Tomoyoshi Ito\authormark{1}}

\address{\authormark{1}Graduate School of Engineering, Chiba University, Yayoi-cho 1-33, Inage-ku, Chiba 263-8522, Japan}
\email{\authormark{*}wangfan@chiba-u.jp} 



\begin{abstract}
A controllable energy method, which considers the undersampling issue of the transfer function and valid spectral energy of a source signal, is proposed to implement angular spectrum diffraction calculation in near and far fields. The proposed method provides an optimized frequency boundary $f_{CE}$ within which it always keeps controllable energy to be diffracted. The controllable energy angular spectrum method significantly reduces the number of samples while having the same accuracy as previous angular spectrum methods, implying a higher calculation efficiency. The new perspective of analyzing spectral energy is shown to improve the performance of relevant diffraction calculations.
\end{abstract}

\section{Introduction}
Scalar diffraction theory plays a pivotal role in optical information processing by providing the fundamental computational method of light wave propagation, called the Rayleigh–Sommerfeld diffraction formula \cite{goodman2005}. In numerical calculations, the Rayleigh–Sommerfeld method is generally classified into three for implementation \cite{mendlovic1997,shen2006,voelz2009}: the single Fourier transform, three Fourier transform based on the impulse response, also known as the convolution-based method, and double Fourier transform based on a transfer function, also known as the angular spectrum (AS) method. The single Fourier transform and three Fourier transform (or convolution-based) methods are based on the Fresnel approximation, whereas the double Fourier transform (or AS) method is strictly derived from the Rayleigh–Sommerfeld formula. The single Fourier transform method produces a target plane with a different size from the source plane because the sampling interval changes with the diffractive distance, whereas the convolution-based and AS methods maintain the same size of two parallel planes. In most cases, we expect that diffraction calculations yield a target plane of the same size as the source plane. Consequently, the AS method has become significant in the field of numerical diffraction computation.\par 

However, the AS method has a well-known limitation that produces considerable aliasing distortion due to undersampling at long-distance diffraction, which has been revealed in many studies \cite{voelz2009, matsushima2009, yu2012, kim2014, zhang2020}. Over the past decade, scholars have proposed their own approaches on how to enable the AS method to be performed over a wide diffraction range, such as the band-limited AS method by Matsushima \cite{matsushima2009}, wide-window AS method by Yu \cite{yu2012}, wide-range AS method by Kim \cite{kim2014}, band-extended AS method \cite{zhang2020} and adaptive sampling AS method \cite{zhang2020_2} by Zhang. In addition, by drawing on the AS diffraction method at long-distance, scholars have further developed methods applied in the off-axis destination plane\cite{matsushima2010shift,zhang2021}, oblique illumination \cite{guo2014} and the polygon-based holograms of large size \cite{tsuchiyama2017}. Therefore, an accurate and effective AS method is required to utilize diffraction calculations widely.\par

\begin{figure}[htbp]
\centering
\includegraphics[width=\linewidth]{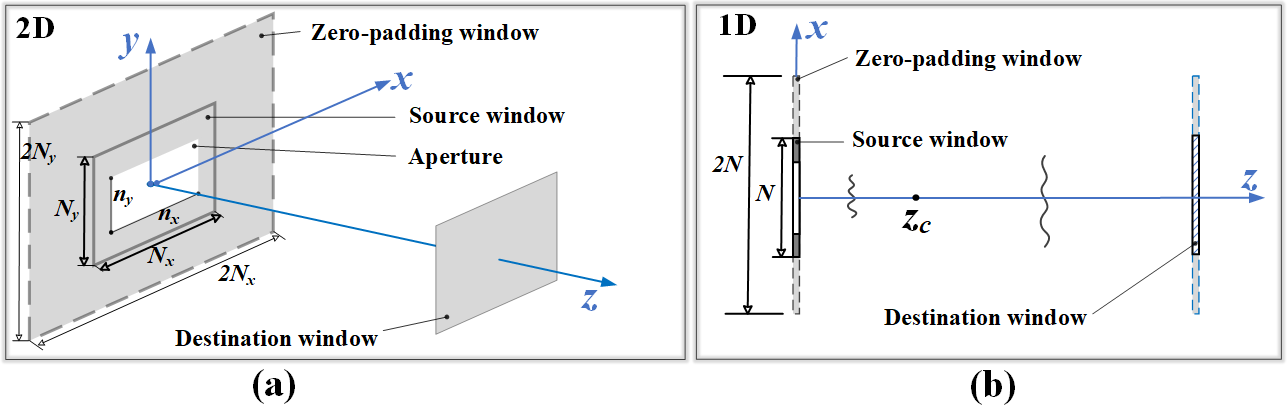}
\caption{Schematic of a plane (a) and a 1D aperture (b) diffraction .}
\label{fig:diffr}
\end{figure}

\section{Sampling analysis of previous AS methods}
Figure. \ref{fig:diffr}(a) shows a diffraction propagation scheme from the source plane to the target plane in the (x,y,z) system. Mathematically, this can be described using the AS method as follows:
\begin{equation} \label{eq:ASdiffr}
\begin{split}
    u(x,y)&=\mathcal{F}^{-1}\left\{A(f_x,f_y) \cdot H(f_x,f_y)\right\}\\
    A(f_x,f_y)&=\mathcal{F}[u_0(x,y)]\,,
\end{split}
\end{equation}
where $u_0(x,y)$ and $u(x,y)$ are the complex amplitude distribution in the source and destination window sampled by $N_x\times N_y$, respectively, the operators $\mathcal{F}$ and $\mathcal{F}^{-1}$ are the fast Fourier transform (FFT) and inverse FFT, respectively. $A(f_x,f_y)$ and $H(f_x,f_y)$ are also known as angular spectrum and transfer function, respectively. $H(f_x,f_y)$ is given by
\begin{equation}
\label{eq:TF}
    H(f_x,f_y)=\exp\left(jkz\sqrt{1-(\lambda f_x)^2-(\lambda f_y)^2}\right)\,,
\end{equation}
where $\lambda$ is the wavelength, $k=2\pi/\lambda$, $j=\sqrt{-1}$, $z$ is the propagation distance, and $(f_x,f_y)$ is the frequency coordinates. In the numerical computation, a considerable error, resulting from the circular convolution in Eq. (\ref{eq:ASdiffr}), can be avoided by doubling the number of samples of the source window and padding the outer area with zeros, as shown the zero-padding window of Fig.  \ref{fig:diffr}(a), which the reason of zero-padding has been widely discussed in \cite{voelz2009,matsushima2009,zhang2020_2}. \par

For simplicity, the following discussion is based on the one-dimensional (1D) case, as shown in Fig. \ref{fig:diffr}(b). From Eq.(\ref{eq:TF}), the phase of transfer function, defined as $\phi(f_x)=kz\sqrt{1-(\lambda f_x)^2}$ in 1D, increases as $z$ gets farther and $f_x$ gets greater. For satisfying the sampling theorem, $\phi(f_x)$ is restricted to
\begin{equation}
\label{eq:shannon}
1/2\pi|\partial \phi/\partial f_x|_{max}\leq 1/2\Delta_f\,.
\end{equation}
$\Delta_f=1/2N\Delta_x$ is the frequency sampling pitch, where $\Delta_x$ is the spatial sampling pitch, and $N$ is the number of samples of the source window, and $2N$ is the number of samples after double zero-padding. A critical distance $z_c=2N\Delta_x^2/\lambda$ was clearly given in  \cite{voelz2009,zhang2020}, which implies that Eq.(\ref{eq:TF}) will be undersampled for greater than $z_c$. \par
\begin{figure}[htbp]
\centering
\includegraphics[width=0.8\linewidth]{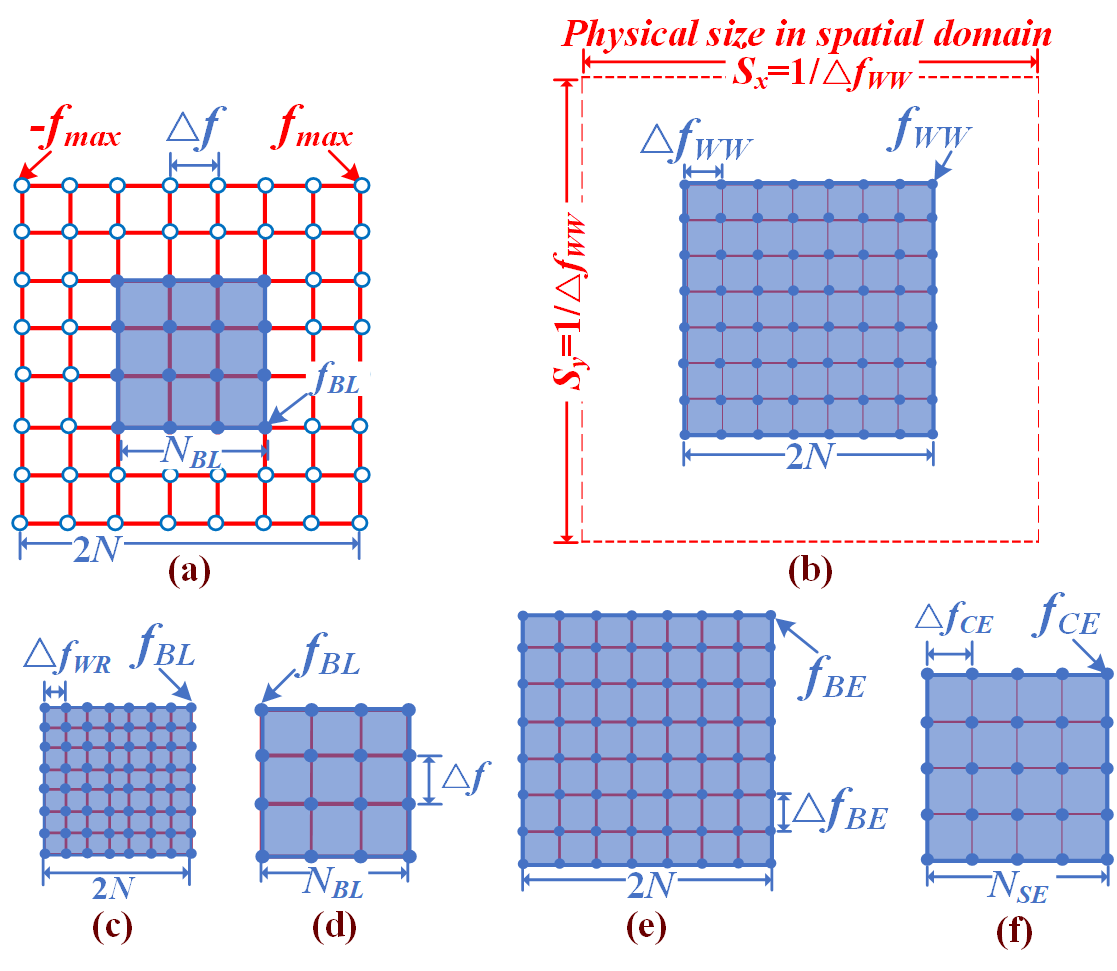}
\caption{An overview of sampling strategy for all AS methods in 2D. The blue shade area is the spectral bandwidth, and blue dots are sampling dots used for calculation. (a) Band-limited AS method, hollow dots mean padding zeros outside the blue area, (b) Wide-window AS method, (c) Wide-range AS method, (d) Adaptive-sampling AS method, (e) Band-extended AS method, (f) The proposed controllable-energy AS method.}
\label{fig:ASMs}
\end{figure}

Now, we briefly review the development of various long-distance AS methods. The band-limited AS method \cite{matsushima2009} presented the first strict derivation of Eq. (\ref{eq:shannon}) that the frequency boundary without undersampling, denoted by $f_{BL}$, depends on
\begin{equation}\label{eq:f_BL}
    f_{BL}=N\Delta_x/z\lambda\,.
\end{equation}
The core of the band-limited AS method is to let $H(f_x)=0$ for everywhere $f_x>f_{BL}$ without changing the number of samples and sampling interval. To intuitively illustrate, Fig. \ref{fig:ASMs} shows a two-dimensional (2D) frequency domain sampling profile for a square window ($2Nx=2Ny=2N$). In Fig. \ref{fig:ASMs}(a), $f_{max}=1/2\Delta_x$ is the maximum frequency that can be recorded by the zero-padding plane, which always satisfies $f_{BL}<f_{max}$ when $z>z_c$. At a very long distance, the calculation results of the band-limited AS method will gradually fail because the number of valid frequency samples,
\begin{equation}\label{eq:N_BL}
N_{BL}=2f_{BL}/\Delta_f=4N\Delta_x f_{BL}=4(N\Delta_x)^2/z\lambda, 
\end{equation}
becomes very small. The wild-window AS method \cite{yu2012} forces the valid number of samples to be $2N$, which means frequency sampling interval will change, denoted by $\Delta_{f_{WW}}$. The authors stated that the width of the calculation window size $S_x\times S_y$ increased with $z$, as shown in the red dotted line region of Fig. \ref{fig:ASMs}(b). The advantage of the wild-window AS method over the band-limited AS method is that the frequency boundary is extended from $f_{BL}$ to $f_{WW}$. However, the wild window AS method requires two convolution operations because FFT is unavailable after changing the frequency sampling interval to be $\Delta_{ f_{WW}}$, which is quite time-consuming \cite{yu2012,yu2012band}. The wild-range AS method \cite{kim2014} and adaptive-sampling AS method \cite{zhang2020_2} follow the valid frequency region given in Eq.(\ref{eq:f_BL}), $f_{BL}$. However, the difference is that the former forces the number of samples to be $2N$ by a new interval $\Delta_{f_{WR}}$, as shown in Fig. \ref{fig:ASMs}(c), whereas the latter holds the original sampling interval $\Delta_f$ constant and uses only the valid number of samples $N_{BL}$ to calculate, as shown in Fig. \ref{fig:ASMs}(d). Owing to the change in sampling interval (in the wild-range AS method ) and the change in sampling number (in the adaptive-sampling AS method), both employ the nonuniform-FFT (NUFFT) technology \cite{NUFFT12,NUFFT3} instead of FFT, and their computational efficiency is better than that of the wild-window AS method but their calculational quality is similar to that of the band-limited AS method. The band-extended AS method \cite{zhang2020} significantly improves on all previous methods and gives a comprehensive analysis of the boundary frequency  
\begin{equation}
\label{eq:f_BE}
    f_{BE}=\sqrt{N/2\lambda z} \,,
\end{equation}
which is greater than $f_{BL}$, as shown in Fig. \ref{fig:ASMs}(e). In fact, the band-extended AS method provides the same boundary as the wild-window AS method, i.e., $f_{BE}=f_{WW}$ and $\Delta_{f_{BE}}=\Delta_{f_{WW}}$, although the wild-window AS method is based on the concept of widening window size. However, the most significant improvement of the band-extended AS method is the use of NUFFT under the frequency region of $f_{BE}$, which makes calculations more economical than in the wild-window AS method and makes results more accurate than in the band-limited, wild-range, and adaptive-sampling AS methods. Table \ref{tab1:methods} summaries the differences between these AS methods, where the texts highlighted in red means to keep parameters constant, whereas the texts highlighted in green mean to adjust parameter as the distance changes. The wild-window and band-extended AS methods have better quality as distance increases due to their wide band, whereas the adaptive-sampling AS method has the lowest workload due to the smallest number of samples.\par
\begin{table}[htbp]
\centering
\begin{threeparttable}[b]
\caption{\bf \centering The list of comparisons between all AS methods.}
\begin{tabular}{lccccc}
\hline
Methods\tnote{\textbf{1}}  & Boundary & Number & Interval  & Load   & Quality\tnote{\textbf{2}} \\
\hline
Band-limited AS\cite{matsushima2009}    & $f_{BL}$$\cdots f_{max}$\tnote{\textbf{3}} & \textcolor{red}{$2N$}   & \textcolor{red}{$\Delta_f$}            & Middle &   Worse  \\
Wild-window AS\cite{yu2012}    & $f_{WW}$   &\textcolor{red}{$2N$}     & \textcolor{green}{$\Delta_{f_{WW}}$}    &High & Better\\
Wild-range AS\cite{kim2014}   &$f_{BL}$    & \textcolor{red}{$2N$}    & \textcolor{green}{$\Delta_{f_{WR}}$}    & Middle &  Worse\\
Adaptive-sampling AS\cite{zhang2020_2}    &$f_{BL}$    & \textcolor{green}{$N_{BL}$}&\textcolor{red}{$\Delta_f$}          & Low    & Worse  \\
Band-extended AS\cite{zhang2020}     &$f_{BE}$& \textcolor{red}{$2N$}&\textcolor{green}{$\Delta_{f_{BE}}$}&Middle &Better\\
Controllable-energy AS\ \tnote{\textbf{4}} & $f_{CE}$&\textcolor{green}{$N_{CE}$}& \textcolor{green}{$\Delta_{f_{CE}}$}  & Low & Better\\

\hline
\end{tabular}
\begin{tablenotes}
\footnotesize
    \item[\textbf{1}] The first two methods are implemented by FFT and two convolutions, whereas the last four are implemented by NUFFT.
  \item[\textbf{2}] Change in quality with the longer distance.
  \item[\textbf{3}] Boundary for computing is $f_{max}$ but the valid one is $f_{BL}$, zero-padding between $f_{BL}$ and $f_{max}$.
  \item[\textbf{4}] The proposed method that keeps spectral energy proportion constant.
\end{tablenotes} 
  \label{tab1:methods}
  \end{threeparttable}
\end{table}

\section{The proposed method}
\begin{figure}[htbp] 
\centering
\includegraphics[width=\linewidth]{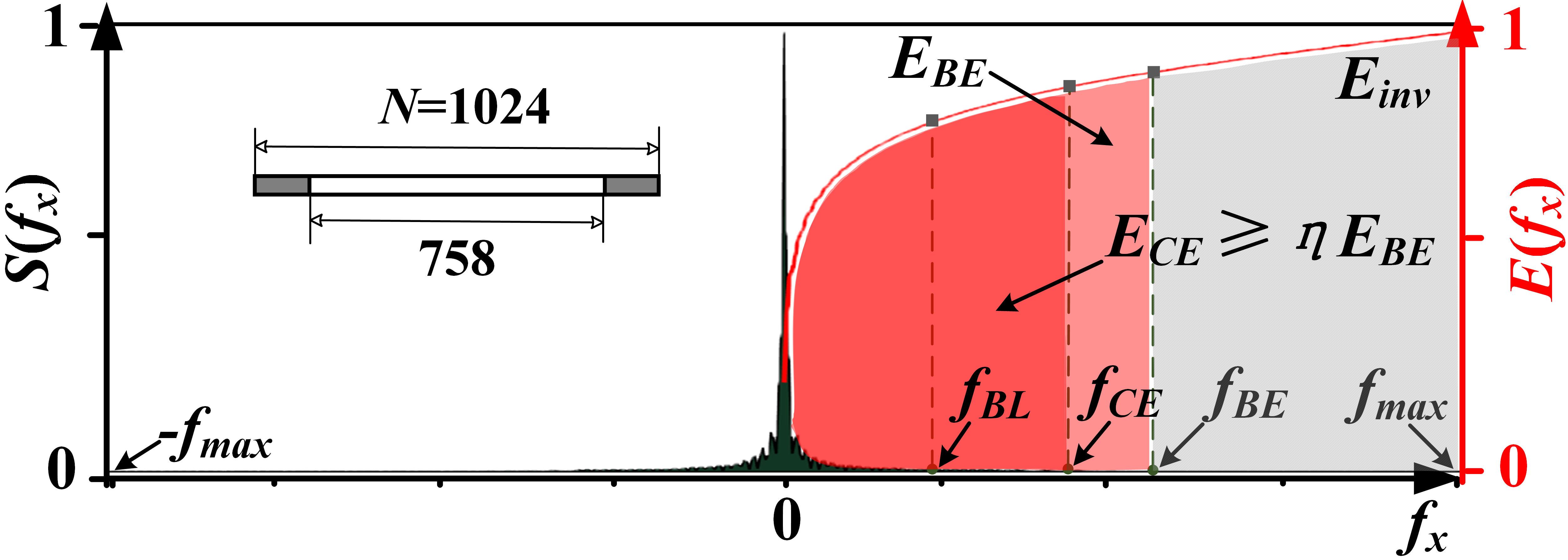}
\caption{Distribution of normalized energy spectral density $S(f_x)$ (left axis) and spectral energy $E(f_x)$ (right axis) with respect to the frequency of a source window, where the number of samples is $N=1024$ and aperture sample is $758$.}
\label{fig:ESD}
\end{figure}

As discussed above, the abovementioned AS methods emphasized the limitation of $H(f_x,f_y)$ in Eq.(\ref{eq:TF}) in the undersampling problem of long-distance and derived the boundary frequency based on Eq. (\ref{eq:shannon}). However, to date, the effect of $A(f_x,f_y)$ in Eq. (\ref{eq:ASdiffr}) has not been closely examined. We argue that, in addition to the transfer function, spectral energy plays a paramount role in diffraction. Thus, in this study, we propose a synthetic adaptive sampling method on the premise that renders the effective energy proportion of spectral $A(f_x,f_y)$ unchanged, named controllable-energy AS method, which has a frequency boundary $f_{CE}$ always between $f_{BL}$ and $f_{BE}$, as shown in Fig. \ref{fig:ESD}. \par

For a 1D aperiodic signal $u_0(x)$, its spectrum is $A(f_x)$. The energy spectral density is define as $S(f_x)=|A(f_x)|^2$, and Parseval's theorem states that the signal energy in the spatial domain can be expressed by the energy in the frequency domain, as follows:
\begin{equation} \label{eq:ESD}
   \int_{-\infty}^{\infty}|u_0(x)|^2\mathrm{d}x= E(f_x)=2\int_0^{f_x} S(f_x)\mathrm{d} f_x ,\qquad 0\leq f_x\leq f_{max} 
\end{equation}
which represents the distribution of signal energy as a function of frequency \cite{ESD}. Notably, the double integral in Eq. (\ref{eq:ESD}) is because the spectrum is normally symmetrically distributed around the zero-frequency.\par

Figure \ref{fig:ESD} is an exemplary spectrum distribution of a simple aperture with 758 samples (the entire source window has 1024 samples) obtained directly by FFT. The black and red curves represent the energy spectral density and spectral energy, respectively, which both indicate the energy is mainly concentrated in the low-frequency components. Although the source signal contains the band $\left[0,f_{max}\right]$, a considerable amount of high-frequency energy is invalid due to the undersampling issue of the transfer function, e.g., the band $\left[f_{BE}, f_{max}\right]$ in the band-extended AS method, $E_{inv}$, as shown in the gray shaded area in Fig. \ref{fig:ESD}. Even within the optimized frequency boundary $f_{BE}$, there are still some redundant sampling points since the high frequencies extremely close to $f_{BE}$ have generally little energy and contribute little to the diffractive results. \par 

From Eq. (\ref{eq:ESD}), we define the spectral energy within $f_{BE}$ as $E_{BE}=2\int_0^{f_{BE}} S(f_x)\mathrm{d} f_x$ and the spectral energy within the proposed frequency boundary $f_{CE}$ as $E_{CE}=2\int_0^{f_{CE}} S(f_x)\mathrm{d} f_x$; therefore, we introduce a user-defined factor $\eta$ to determine the proportion of $E_{CE}$ to $E_{BE}$, which can be mathematically expressed by
\begin{equation} \label{eq:Ese}
    E_{CE}\ge \eta E_{BE},\qquad 0\leq \eta\leq 1\,.
\end{equation}
This equation is the kernel of the proposed method, which suggests that the spectral energy the proposed method used is always proportional to that used by the band-extended AS method. Thus, we mainly seek the frequency boundary $f_{CE}$ that just satisfies Eq. (\ref{eq:Ese}) from $0$ to $f_{max}$.\par

To quickly find $f_{CE}$, we can search from $f_{BL}$ given in Eq. (\ref{eq:f_BL}), because the energy within $f_{BL}$ is always less than $E_{BE}$. Algorithm \ref{alg} provides the pseudocode to search the appropriate value of $f_{CE}$, where $\lceil\cdot\rceil$ rounds numbers to upper integers of their argument. Theoretically, $\eta$ can be any value between 0 and 1, which depends on the quality of results a user wants. Empirically, the quality of results using $f_{CE}$ as boundary are very consistent with those of using $f_{BE}$ when $\eta>0.9$. However, a major advantage of the proposed controllable-energy AS method is that the number of samples is well below that of the band-extended AS method, but their computational accuracies are the same, which implies that the computational efficiency of the proposed controllable-energy AS method is able to well above that of the band-extended AS method.\par
\begin{algorithm}
\caption{Searching for $f_{CE}$ based on the given value of $\eta$.} 
\label{alg}
\begin{algorithmic}[1]
\State $\eta \gets 0.995$ \Comment{User-defined value (between 0 and 1)}
\State $E_{BE}\gets 2\times\sum\limits_{m=0}^{\lceil f_{BE}/\Delta_f\rceil}|A(m\Delta_f)|^2\cdot \Delta_f$
\State $j \gets \lceil f_{BL}/\Delta_{f}\rceil$
\State $E_{CE}\gets 2\times\sum\limits_{m=0}^{j}|A(m\Delta_{f})|^2\cdot\Delta_{f}$
\While{$E_{CE}\leq \eta E_{BE}$} 
\State $j\gets j+1$
\State $E_{CE}\gets E_{CE}+2|A(j\Delta_{f})|^2\cdot\Delta_{f}$
\EndWhile
\State \textbf{return} $f_{CE}\gets j\Delta_{f}$
\end{algorithmic}
\end{algorithm}

We now analyze the number of samples of the proposed controllable-energy AS method. Under the constraint of the boundary frequency $f_{CE}$ searched by Eq. (\ref{eq:Ese}), we define a new sampling interval of frequency as $\Delta_{f_{CE}}=2f_{CE}/N_{CE}$, where $N_{CE}$ is the number of samples of the controllable-energy AS method, as shown in Fig. \ref{fig:ASMs}(f). As a result, Eq. (\ref{eq:shannon}) becomes
\begin{equation}\label{eq:N_SE1}
\begin{split}
    \lambda \frac{z f_{CE}}{\sqrt{1-(\lambda f_{CE})^2}}\leq \frac{N_{CE}}{4f_{CE}}\,,\\
    \Rightarrow{N_{CE}\ge \frac{4\lambda zf^2_{CE}}{\sqrt{1-(\lambda f_{CE})^2}}}   \,.
\end{split}
\end{equation}
Considering that $(\lambda f_{CE})^2\ll 1$ holds in most of cases, and taking the minimum value of $N_{CE}$, then 
\begin{equation}\label{eq:N_SE2}
    N_{CE}\approx\left \lceil 4\lambda zf^2_{CE} \right\rceil\,.
\end{equation}
As mentioned above, there is $N\Delta_x/z\lambda=f_{BL}\leq f_{CE}\leq f_{BE}=\sqrt{N/2\lambda z}$, which indicates that $N_{BL}\leq N_{CE}\leq 2N$ always exists. Therefore, the effective angular spectrum of the source window with the sampling interval $\Delta_{f_{CE}}$ and the number of samples $N_{CE}$ can be discretely calculated as follows:
\begin{equation} \label{eq:A(f_SE)}
A_{CE}(f_x)={\rm NUFFT_{3}} \{u_0(x)\},\  
\end{equation}
where $x=i\Delta_x, i\in[0,N)$ and $f_x=m\Delta_{f_{CE}},m\in[0,N_{CE})$. The operation ${\rm NUFFT_3}\{\cdot\}$ denotes NUFFT, which has been effectively used by many scholars to overcome the limitation when either sampling interval or number of samples of the input signal does not match those of the spectrum \cite{kim2014,zhang2020,zhang2020_2,shimo2012scaled,shimo2013NUfft,chang2014nufft}. Herein, we adopt the type-3 NUFFT proposed by Lee and Greengard \cite{NUFFT3}. Based on Eq. (\ref{eq:A(f_SE)}), the diffraction field by the proposed controllable-energy AS method can be obtained as follows
\begin{equation}
    u(x)={\rm NUFFT_3}^{-1}\{A_{CE}(f_x)\cdot H(f_x)\}\,,
\end{equation}
where the superscript $^{-1}$ denotes inverse NUFFT. The proposed controllable-energy AS method can be readily expanded to the 2D case.\par

\section{Results}
\begin{figure}[htbp] 
\centering
\includegraphics[width=\linewidth]{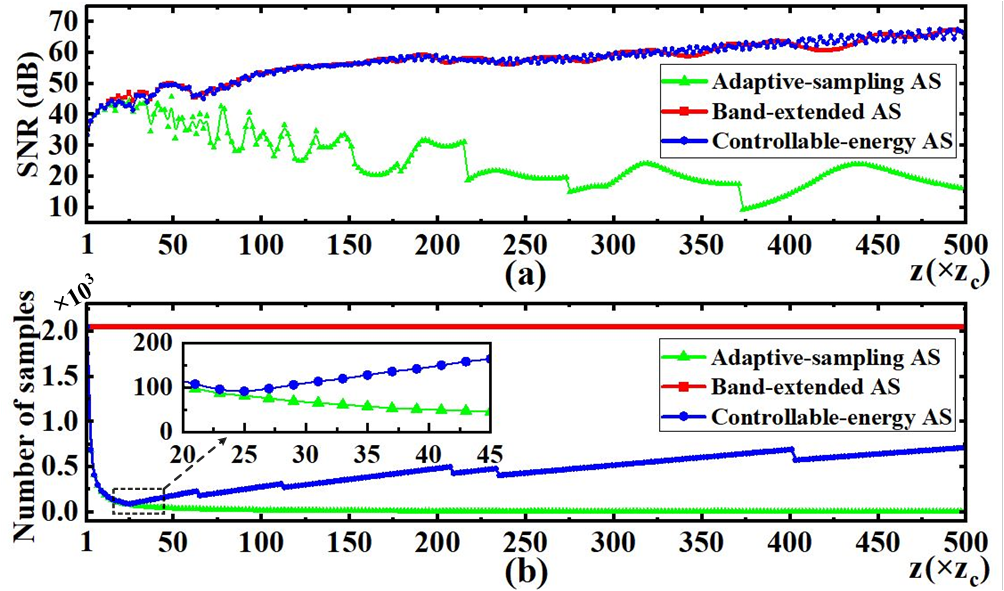}
\caption{Results comparison for the adaptive-sampling AS, band-extended AS and proposed controllable-energy AS methods ($\eta$=$0.995$) between $z=z_c$ and $z=500z_c$. (a) Comparison of the SNR of amplitude values, and (b) comparison of the number of samples used for calculation.}
\label{fig:result1}
\end{figure}

As a comparison, we use the signal-to-noise ratio (SNR) as the evaluation criterion in this manuscript, which is widely used in the previous methods\cite{matsushima2009,yu2012,kim2014,zhang2020,zhang2020_2}. Considering the amplitude of diffraction field by the analytical Rayleigh–Sommerfeld diffraction integral as reference data and the amplitude of diffraction field by the adaptive-sampling AS, band-extended AS, and proposed controllable-energy AS methods as evaluation data, Fig. \ref{fig:result1} shows the evaluation results with the parameters of wavelength $\lambda=532{\rm nm}$, pixel pitch $\Delta_x=1\upmu\rm{ m}$, source window samples $N=1024$, and aperture samples $758$, propagating from $z_c(=3.8{\rm mm})$ to $500z_c$. \par 

From Fig. \ref{fig:result1}(a), under $\eta=0.995$, the results of the proposed controllable-energy AS method and the band-extended AS method performed well as the distance increased, and both were highly consistent, whereas the adaptive-sampling AS method results worsened at longer distances due to the sharp shrinking of the boundary frequency $f_{BL}$, see Eq. (\ref{eq:f_BL}). Moreover, Fig. \ref{fig:result1}(b) shows that the number of samples of the proposed controllable-energy AS method significantly reduced, even when holding $99.5\%$ energy of $E_{BE}$, whereas the band-extended AS method always maintained $2N$ samples. In the subgraph of Fig. \ref{fig:result1}(b), with the current parameters, the proposed controllable-energy AS method had the same number of samples as the adaptive-sampling AS method within around $25z_c$, which indicated they had the same boundary frequency because the spectral energy of the adaptive-sampling AS method, $E_{AS}$, was greater than $0.995E_{BE}$. At a longer distance than $25z_c$, the adaptive-sampling AS method lost more energy so that $\textbf{while}\mathbf{\cdots}\textbf{do}$ loop in Algorithm \ref{alg} was launched to search the proposed frequency boundary $f_{CE}$. To keep a steady proportion of energy, $N_{CE}$ needs to automatically increase with distance but not more than $2N$. How fast $N_{CE}$ increases depends on the value of $\eta$ and how much high-frequency energy the source signal has. \par
\begin{figure}[htbp] 
\centering
\includegraphics[width=0.6\linewidth]{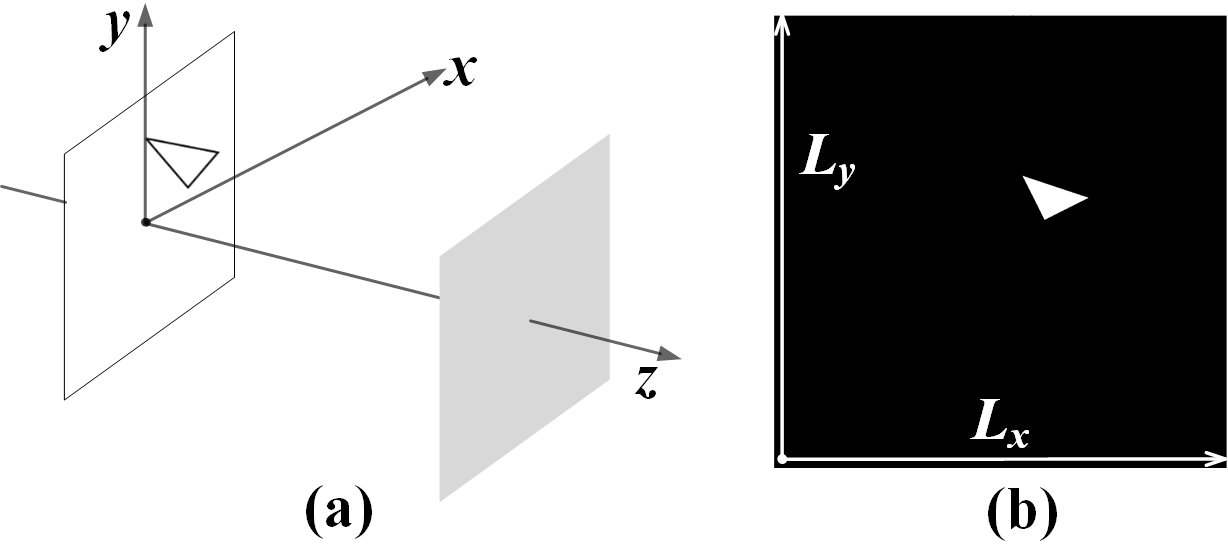}
\caption{An arbitrary triangle with vertexes $(0.05, 0.15, 0)$mm, $(0.1, 0.05, 0)$mm, $(0.2, 0.1, 0)$mm is used for the diffraction calculation. (a) Schematic diagram of the triangle in the three dimension coordinates system. (b) The rasterization of the triangle in the two dimensional plane.}
\label{fig:result2}
\end{figure}

Overall, the comparison of these methods revealed that the proposed controllable-energy AS method had a lighter workload than the band-extended AS method without a loss of computational accuracy. We further implemented the proposed method for an arbitrary triangle, as shown in Fig. \ref{fig:result2}(a). The triangle lies in the plane $z=0$, with the vertexes $(0.05, 0.15, 0)$, $(0.1, 0.05, 0)$, $(0.2, 0.1, 0)$, where the unit is millimeter (mm). Figure \ref{fig:result2}(b) shows the rasterized triangle on the plane with size $L_x\times L_y$, which is parallel to the target plane. The number of pixels and sampling interval were 1024$\times$1024 and 1$\upmu\rm{m}$$\times$$1\upmu\rm{m}$, respectively. \par 

\begin{figure}[htbp] 
\centering
\includegraphics[width=\linewidth]{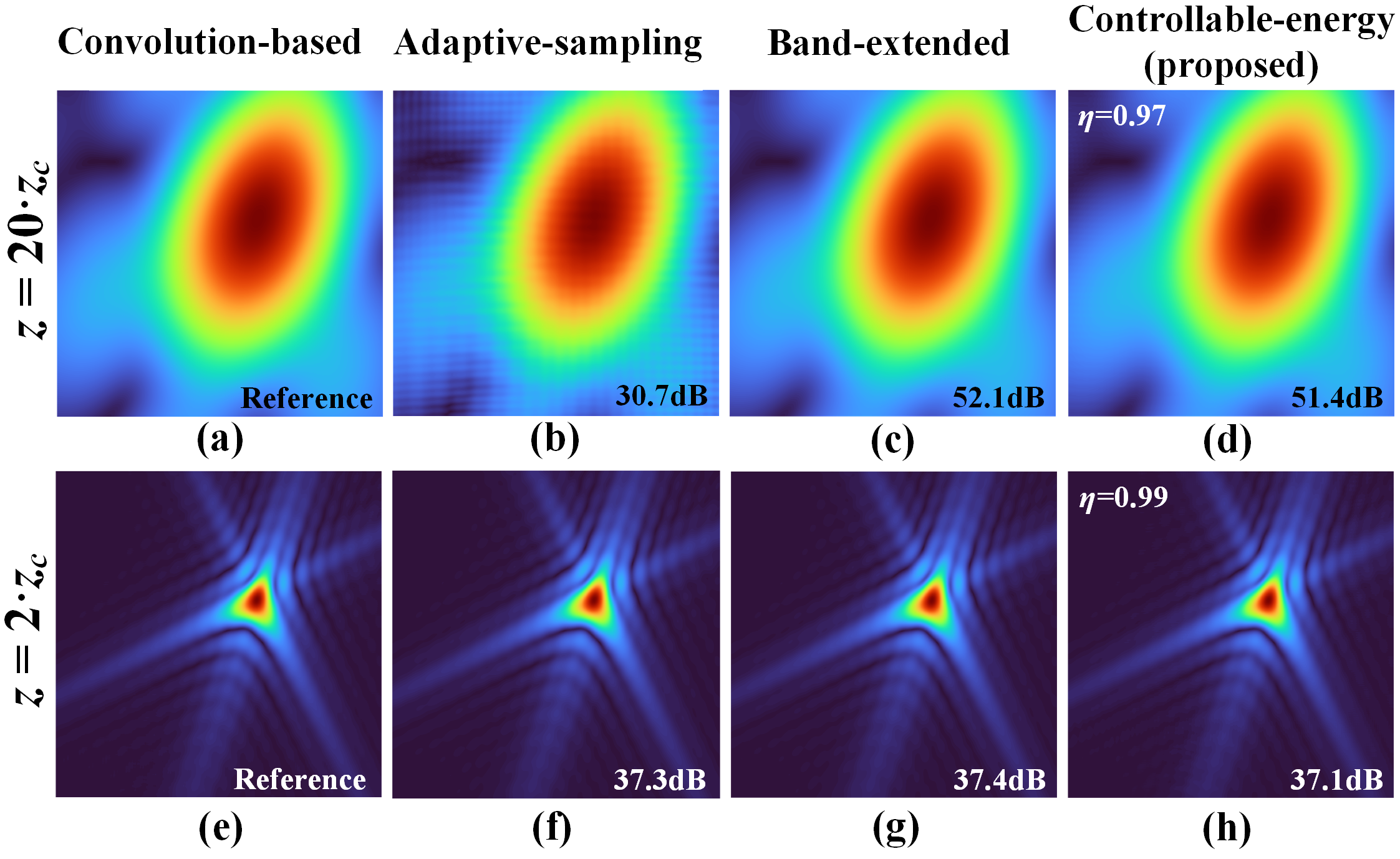}
\caption{ Diffraction results of the triangle obtained by different methods at $z=20z_c$ and $z=2z_c$. The baseline of the comparison is the convolution-based method.}
\label{fig:result3}
\end{figure}

Figures \ref{fig:result3}(a-d) depict diffraction results of the convolution-based (three Fourier transforms) method, adaptive-sampling AS method, band-extended AS method and proposed method with $\eta=0.97$ at the position $z=20z_c$, respectively. For comparing the quality of results, we took the diffraction filed obtained by the convolution-based method as reference. From Fig.\ref{fig:result3}(b), the adaptive-sampling AS method induces noises because of the loss of high frequency information. However, the proposed method offers the same high accuracy as the band-extended AS method, as shown in Figs.\ref{fig:result3}(c) and (d). Under the environment of CPU with AMD Ryzen 5 3600 @3.59GHz, the detailed results are listed in Table \ref{tab: results}. Owing to the reduction in the number of sampling, the elapsed times of the adaptive-sampling AS and the proposed method were 0.86 s and 1.06 s, respectively, whereas that of the band-extended AS method was 4.33 s, i.e., approximately 4 times acceleration by the proposed method. Note that, the elapsed time is not linearly related to the number of sampling because the interpolation required in the NUFFT technology takes a constant expense \cite{NUFFT12,NUFFT3}.\par
\begin{table*}[htbp]
    \centering
        \caption{Comparison of results at different distances and $\eta$.}
    \label{tab: results}
    \renewcommand{\arraystretch}{1.2}
        \begin{tabular}{m{2em}<{\centering}|m{2em}<{\centering}|m{3em}m{6em}<{\centering} m{5em}<{\centering} m{9em}<{\centering}}
    \hline
     z  & $\eta$  &     &   Adaptive-sampling AS& Extended-band AS & Controllable-energy AS (Proposed)  \\
     \hline
     \multirow{3}{2em}{$20 z_c$} & \multirow{3}{2em}{0.97}   & SNR(dB)   & 30.7  & 52.1   & 51.4\\
  \cline{3-6}
                                                                 & & Time(s)  & 0.86  & 4.33  & 1.06 \\
                                                                 \cline{3-6}
                                                                 & & Sampling pixels  &$102\times 102$ & $2048\times 2048$  & $448\times 448$\\
    \hline
    \multirow{3}{2em}{$2 z_c$} & \multirow{3}{2em}{0.99}   & SNR(dB)   & 37.3  & 37.4   & 37.2\\
    \cline{3-6}
                                                               &   & Time(s)  & 1.3  & 4.4  & 1.00 \\
                                                               \cline{3-6}
                                                              &    & Sampling pixels & $683\times 683$ & $2048\times 2048 $  & $406\times 406 $\\ 
    \hline
    \end{tabular}

\end{table*}

As a discussion, it is worth noting that the spectral energy suggested by the proposed method is determined by the value of $\eta$ and the spectral energy of the band-extended method, see Eq. (\ref{eq:Ese}), but this does not exclude a broader application. In Fig. \ref{fig:result3}(d), the factor of $\eta$ can be further reduced depending on the user's requirement for image quality because a processed image is acceptable to human eyes if its peak-SNR is greater than 30 dB \cite{psnr_chen1998}, which implies a smaller number of samples. Moreover, the spectral energy $E_{BE}$ in Eq. (\ref{eq:Ese}) can be replaced with the spectral energy within the frequency boundary $f_{BL}$ at the closer position, because  the accuracy of the adaptive-sampling method and the band-extended method is almost the same within a distance of about $10z_c$. By doing so, Eq. (\ref{eq:Ese}) is rewritten as 
\begin{equation}
       E_{CE}\ge \eta E_{BL},\qquad 0\leq \eta\leq 1\,,
       \label{eq:Ese2}
\end{equation}
where $E_{BL}=2\int_0^{f_{BL}} S(f_x)\mathrm{d} f_x$. This indicates that we can further compress the spectral region of $f_{CE}$. Unsurprisingly, if $f_{CE}<f_{BL}$, the computational effort will be significantly reduced. For some intensive calculation cases in the frequency domain, such as the calculation of the polygon-based hologram \cite{tsuchiyama2017}, the proposed method offers flexibility in choosing the trade-off between efficiency and accuracy. To simply confirm this application, Figs. \ref{fig:result3}(e-h) provide the diffraction results for the triangle in Fig. \ref{fig:result2} at $z=2z_c$. Figure \ref{fig:result3}(h) was obtained based on Eq. (\ref{eq:Ese2}) and $\eta=0.99$. From Table \ref{tab: results}, the number of sampling by the proposed method is smaller than that of the adaptive-sampling AS method with the comparable quality.

\section{Conclusion}
In conclusion, we proposed controllable-energy method to perform diffraction calculations at near to far distances based on the adaptive-sampling AS method \cite{zhang2020_2} and the band-extended AS method \cite{zhang2020}. The proposed method can effectively reduce the number of samples compared with the band-extended AS method while maintaining higher accuracy than the adaptive-sampling AS method. We further confirmed that the proposed method can be flexibly adapted to the user's needs, resulting in more economical computations.\par
This study verified the effectiveness of the proposed controllable-energy AS method in the diffraction calculation. Moreover, significantly, it developed a new perspective of considering the spectrum energy in the diffraction propagation issues rather than just considering the undersampling problem, which may provide some insights for related studies.

\section*{Funding}
This work was partially supported by JSPS KAKENHI Grant Numbers 19H04132 and 19H01097.
\section*{Acknowledgements}
We thank the Ministry of Education, Culture, Sports, Science and Technology (MEXT) of
Japan for their support. And thanks to Dr. Wenhui Zhang and Dr. Haitong Sui for discussions.
\section*{Disclosures}
The authors declare no conflicts of interest.
\section*{Data Availability} 
Data underlying the results presented in this Letter are not publicly available at this time but may be obtained from the authors upon reasonable request.

\bibliography{CE-AS_method}


\end{document}